\documentclass[conference]{IEEEtran}
\IEEEoverridecommandlockouts
\usepackage{cite}
\usepackage{amsmath,amssymb,amsfonts}
\usepackage{algorithm}
\usepackage{algorithmic}
\usepackage{graphicx}
\usepackage{textcomp}
\usepackage{xcolor}
\usepackage{graphicx}
\usepackage{epstopdf}
\usepackage{float}
\usepackage{bm}
\usepackage{subfigure}
\usepackage{diagbox}
\usepackage{multirow}
\usepackage{booktabs}
\def\BibTeX{{\rm B\kern-.05em{\sc i\kern-.025em b}\kern-.08em
		T\kern-.1667em\lower.7ex\hbox{E}\kern-.125emX}}
	
\begin{document}
	
\title{	\LARGE A Parallel Distributed Algorithm for the Power SVD Method}

\author{
	Jiaying Li$^\dag$, \quad Sissi Xiaoxiao Wu$^\dag$,\quad  Qiang Li$^\ddag$ and Anna Scaglione$^\star$\\
	\small
	$^\dag$College of Elec. and Info. Eng., Shenzhen Univ., China, \quad 
	$^\ddag$School of Info. and Comm. Eng., UESTC, China \\
	\small
	$^\star$School of Electrical, Computer and Energy Engineering, Arizona State Univ., Tempe, AZ, USA 
}

\maketitle

\begin{abstract}
In this work, we study how to implement a distributed algorithm for the power method in a parallel manner.  As the existing distributed power method is usually sequentially updating the eigenvectors, it exhibits two obvious disadvantages: 1) when it calculates the $h$th eigenvector, it needs to wait for the results of previous $(h-1)$ eigenvectors, which causes a delay in acquiring all the eigenvalues; 2) when calculating each eigenvector, it needs a certain cost of information exchange within the neighboring nodes for every power iteration, which could be unbearable when the number of eigenvectors or the number of nodes is large. This motivates us to propose a parallel distributed power method, which simultaneously calculates all the eigenvectors at each power iteration to ensure that more information could be exchanged in one shaking-hand of communication. We are particularly interested in the distributed power method for both an eigenvalue decomposition (EVD) and a singular value decomposition (SVD), wherein the distributed process is proceed based on a gossip algorithm. It can be shown that, under the same condition, the communication cost of the gossip-based parallel method is only $1/H$ times of that for the sequential counterpart, where $H$ is the number of eigenvectors we want to compute, while the convergence time and error performance of the proposed parallel method are both comparable to those of its sequential counterpart.
\end{abstract}


\section{Introduction}
Principal component analysis (PCA) is the basis of many machine learning algorithms and array processing algorithms, thereby having a wide range of applications in many fields \cite{bishop2006pattern,jolliffe2016principal,xie2017application,o2016distributed,sun2020principal}. 
%
To conduct a PCA, the key is to perform an eigenvalue decomposition (EVD) or a singular value decomposition (SVD) on the covariance (correlation) matrix of the data/signal to obtain the principal domain ($p$-D) eigenvectors.  In this context, to deal with a batch of data/signal, a classic approach is to perform the so-called {\it power method} \cite{mises1929praktische}, which repeatedly performs linear projections on an initial vector until it converges to the direction of the desired eigenvector.  
The power method and its variants have been widely used in many areas\cite{page1999pagerank,lin2010power,ogbe2017noisy,salvati2019power,de2018accelerated,mai2019noisy}.
Early applications of the power method are mainly focused on the centralized mechanism, wherein data is uploaded to a fusion center and the solution is obtained in a centralized manner. In recent years, due to the latest incarnation of networked technologies from social mobile media to the Internet of Things, data is usually distributively located and exponentially growing to analyze. This backdrop has sparked significant advances over the last decades from centralized algorithms to multi-agent signal processing algorithms. The latter has motivated the scenario setting we study in this work. 

Given that data/signal is stored locally at terminal nodes, the basic idea of decentralized PCA is to let each terminal node utilize its local computational resource 
and the ability of neighborhood communication to find the solutions of PCA in a cooperative way. In this context, a type of distributed power method (DPM) has been widely studied in the literature. For example, the authors in \cite{scaglione2008decentralized} first proposed the DPM to calculate the eigenvectors of the covariance matrix of the batch data in the application of distributed direction of arrival (DDoA) estimation and its convergence is well studied in \cite{suleiman2016performance,wai2017fast}. Ref. \cite{raja2015cloud} used the DPM as a subroutine in the process of collaborative dictionary learning. The DeFW algorithm proposed in \cite{wai2017fast} incorporated a DPM step to reduce the complexity caused by distributed computation over network. \cite{mohammadi2016decentralized} used the DPM to estimate the eigenvalue in a decentralized inference problem for spectrum sensing.  \cite{wu2019decentralized} extended the DPM to the case of calculating the non-Hermitian matrix by using a power SVD method. It is worth noting that, in the above applications, some applications simply need the largest eigenvector/eigenvalue \cite{scaglione2008decentralized,raja2015cloud,wai2017fast}, while others \cite{suleiman2016performance, mohammadi2016decentralized, wu2019decentralized} calculate multiple eigenvectors/eigenvalues. More applications could be found in the review paper \cite{wu2018review}. 

While the aforementioned DPM works well in many applications, it suffers from a serious efficiency problem, as it usually works in a sequential updating manner. That is, at the $h$th iteration, the $h$th eigenvector is calculated based on the final results of previous $(h-1)$ eigenvectors. This is a time-consuming procedure as it always needs to wait for the completion of previous eigenvectors to proceed the next one. In some applications such as passive radar \cite{wu2019decentralized} and spectrum sensing \cite{mohammadi2016decentralized}, an estimate of the eigenvalues, even if it is not absolutely accurate, is still meaningful to the system. This motivates us to find an alternative approach which can calculate all the eigenvalues simultaneously. On the other hand, the sequential power method experiences an expensive communication cost, since at each power iteration the system needs to launch a shaking-hand process between the neighboring nodes to exchange information for the calculation of only one eigenvector. As the shaking-hand process is generally expensive, it could more efficient to design a managing decentralized algorithm, which exchanges more information in one time shaking-hand, so as to save the number of shaking hands on overall, given that a comparable convergence rate and estimation accuracy are achievable compared to the sequential DPM. The parallel power method in this work is just proposed to serve these purposes.

In addition to the above works, it is also worth mentioning a series of works \cite{penna2014decentralized,penna2012decentralized} which utilize the so-called Lanzcos algorithm (LA) to perform a parallel PCA. LA was originally proposed by Lanczos \cite{lanczos1950iteration}, which transforms a symmetric matrix into a symmetric tridiagonal matrix through an orthogonal transformation, while keeping the eigenvalues unchanged. The distributed LA in \cite{saad2003iterative} can generally converge fast, while it has some restrictions: 1) it can only calculate eigenvalues but not eigenvectors; 2) the number of LA iteration is tied with (no larger than) the number of calculated eigenvalues; 3) when the number of LA iteration is very large, there might be repeated roots which might deteriorate the distributed PCA performance; 4) it can only deal with Hermitian matrices via EVD.   
The proposed parallel DPM can overcome the above restrictions and provide a good PCA performance for different applications.
Our simulation results show that the proposed parallel DPM is effective as its performance is comparable with the sequential counterpart, while only $1/H$ times of node-to-node shaking-hand is needed for  message passing, where $H$ is the number of eigenvectors we need to compute. In addition, as the number of iterations in the gossip algorithm increases, its performance can even approach to that of the centralized counterpart. We also show that parallel DPM can outperform distributed LA in terms of estimation accuracy.

{

\section{Problem formulation and Preliminaries}
We consider the scenario that two sets of nodes, set $\mathcal{S}$ and set $\mathcal{R}$ store their received signals or data in the local nodes, denoting by  $\bm{s}(t)=(s_1(t),\ldots,s_{|\mathcal{S}|}(t))^T$ and $\bm{r}(t)=(r_1(t),\ldots,r_{|\mathcal{R}|}(t))^T$. The received signal is considered to be zero-mean over $t$, i.e.:
$
 {\mathbb E}\left[{\bm s}(t) \right]= {\mathbb E}\left[{\bm r}(t) \right] =0.
$
Targeting at this model, we consider the PCA problems involving finding solutions of EVD on either set $\mathcal{S}$ or $\mathcal{R}$, and finding the solutions of SVD on both sets, as follows.

\subsection{The PCA problem}
{\it \textbf {1)~The covariance matrix of $\bm{s}(t)$ ($\bm{r}(t)$):}} As the signal is zero-mean, the solutions of PCA in this case can be recast to EVD for the Hermitian matrices
$
{\mathbb E}[{\bm s}(t){{{\bm s}}^{H}}(t)]=\bm{R_{ss}}$ or
$
{\mathbb E}[{\bm r}(t){{{\bm r}}^{H}}(t)]=\bm{R_{rr}}.
$
This PCA problem can be widely seen in many applications, such as channel estimation in massive-MIMO systems, recommendation system in data mining, target detection and (direction-of-arrival) DOA estimation in radar system and so on \cite{xu2016downlink,symeonidis2016matrix,dai2019extended,santamaria2017passive}.  

{\it \textbf {2)~The correlation matrix of $\bm{s}(t)$ and $\bm{r}(t)$:}}
Another type of PCA is the SVD for the correlation matrix
$
{\mathbb E}[{\bm s}(t){{{\bm r}}^{H}}(t)]=\bm{R_{sr}}$ or
${\mathbb E}[{\bm r}(t){{{\bm s}}^{H}}(t)]=\bm{R_{rs}}$.
A typical application is the passive radar system wherein the surveillance sensor set and reference sensor set work together to detect objects by processing reflections from non-cooperative sources of illumination in the environment\cite{wu2019decentralized,santamaria2017passive}.
Note that SVD is a more generalized PCA problem including EVD as a special case.

\subsection{Decentralization and average consensus gossip (ACG)}

To perform a decentralzied PCA, 
we assume that all the nodes in set $\mathcal{S}$ ($\mathcal{R}$) are locally distributed and the  communication network underlying $\mathcal{S}$ ($\mathcal{R}$) is a directed and fully-connected graph $G=(\mathcal{S},E_{\mathcal{S}})$ ($G=(\mathcal{R},E_{\mathcal{R}})$). 
That is, each node has its own observation of the data/signal, i.e., 
$s_{i}(t)$, $r_{j}(t) \in \mathbb{C}$ where $i \in \mathcal{S}$ and $ j \in \mathcal{R}$. Our problem is how to decentralize the computation of the PCA process into each node. 
The main ingredient is the so-called average consensus gossip (ACG) algorithm. Therein, the most important part is to decentralize the computing of the inner product ${\bm a}^{H}{\bm b}$, wherein each node has one entry of ${\bm a}$ (${\bm b}$), i.e., node $i$ stores $a_i$ and $b_i$.  The inner product can be distributively calculated as follows: 1) node $i$ initializes its message to its neighbors with ${{Z}_i}[0]={a}_i^* {b}_i$; 2) node $i$ updates the message conveying from other near neighbors as follow:
\begin{align}\notag
{{Z}_{i}}[k+1]={{W}_{ii}}{{ Z}_{i}}[k]+\sum\nolimits_{j\in {{\mathcal{N}}_{i}}}{{{W}_{ij}}{{Z}_{j}}[k]}, ~k=1,...,K,
\end{align}
where $K$ denotes the total gossip iteration, $\mathcal {N}_{i}$ is the set of neighbors belonging to the node $i$ and ${{\bm W}}$ is the weighting matrix of graph $G$ whose weights ${{W}_{ij}}$ form a convex combination of the node $i$ and its neighbors values. Given that the weights $\bm W$ is doubly stochastic and symmetric (i.e., $\bm{W1}=\bm{1}$ and ${\bm 1}^{T}{\bm W}=\bm{1}^T$) and if the second largest eigenvalue of $\bm W$, ${{\lambda }_{2}}(\bm W)<1$, for all $i$, then ${{Z}_{i}}[k]$ will converge to the average value $\frac{1}{n}{{\bm a}^{H}}{\bm b}$ when $k\to \infty$ with $n$ being the dimension of  vectors ${\bm a}$ and ${\bm b}$. As introduced in  \cite{wu2019decentralized}, an important
property of ACG is that the magnitude of the approximation error decreases exponentially with the number of iteration $k$; for a finite number $k$, the result of average consensus $n{Z}_{i}[k]$ is the approximation of the inner product term  ${{\bm a}^{H}}{\bm b}$, and we denote this result as ${AC}_{i}({{\bm a}^{H}}{\bm b})$, where ${AC}_{i}(\cdot)$ denotes the ACG operator.

\section{The sequential distributed power method } 
The power SVD method could be formulated by solving the following optimization problem:
\begin{align}
	&\max_{{\bm u}, {\bm v} \in {\mathbb R}^n } ({\bm u}^T~{\bm v}^T) \mathbb{E}\left(\begin{matrix} \alpha {\bm I} & {\bm s}(t){{{\bm r}}^{H}}(t) \\ {\bm r}(t){{{\bm s}}^{H}}(t) &\alpha {\bm I} \end{matrix} \right)\left({\bm u} \atop {\bm v}\right) \\\notag
	&~~\mbox{s.~t.~} \quad \|{\bm u}\|=1, \quad \|{\bm v}\|=1,
\end{align}
where ${\bm u}$ and ${\bm v}$ are the left principal and right principal eigenvetors of the matrix $\bm{R_{sr}}$ and $0<\alpha<1$ is a prescribed parameter to maintain the positive-definiteness. In view of this, a centralized power SVD process can be written as
\begin{align}
	\left({\bm u}^{(\ell+1)}\atop {\bm v}^{(\ell+1)}\right) 
	=&\frac 1 T\sum_{t=0}^{T-1}\left(\bm{s}(t)(\bm{r}^H(t){\bm v}^{(\ell)})+ \alpha {\bm u}^{(\ell)} \atop  \bm{r}(t)(\bm{s}^H(t){\bm u}^{(\ell)})+ \alpha {\bm v}^{(\ell)}\right), 
\end{align}
where $\ell$ denotes the power iteration and the final eigenvectors should be normalized by $\left(\hat{\bm u}; \hat{\bm v}\right) = \left({\bm u}^{\ell^\star}/\|{\bm u}^{\ell^\star}\|; {\bm v}^{\ell^\star}/\|{\bm v}^{\ell^\star}\|\right)$ with $\ell^\star$ being the sufficient number of power iteration.
Our previous work on sequential distributed algorithm boils down into two main ingredients \cite{wu2019decentralized}: the power method and the ACG algorithm. 
That is, at the iteration of calculating the $h$th singular vectors, each node in the network locally calculates the top $m$ ($m < h$) singular values:
\begin{align}
	\hat{\sigma}_{im}^2=&
	\sum_{t=0}^{T-1} s_i(t)A{{C}_{i}}({{\bm r}^{H}}(t){{\hat{\bm v}}_{m}})/(T\hat{u}_{im}^*), \quad i \in \mathcal{S}, \label{decen_svd_u_eigenvalue}\\\label{decen_svd_v_eigenvalue}
	\hat{\sigma}_{jm}^2=&
	\sum_{t=0}^{T-1} r_j(t)
	A{{C}_{j}}({{\bm s}^{H}}(t){{\hat{\bm u}}_{m}})/(T\hat{v}_{jm}^*), \quad j \in \mathcal{R}.
\end{align}

Armed with these results, each node can further calculate an approximated entries of left and right eigenvectors ${\bm u}_{h}$ and ${\bm v}_{h}$ in a decentralized fashion by:
\begin{align}
u_{ih}^{(\ell+1)}&=\frac{1}{T}
\sum_{t=0}^{T-1}s_i(t)
A{{C}_{i}}({{\bm r}^{H}}(t){\bm v}_{h}^{(\ell)}) + \alpha {u}_{ih}^{(\ell)}\label{decen_svd_u}\\\notag
&-\frac{1}{T}\sum_{m=1}^{h-1}\sum_{t=0}^{T-1}{{\hat\sigma_{im}}}^2
\hat{u}_{im}
A{{C}_{i}}({\hat{{\bm v}}_m^{H}}{\bm v}_{h}^{(\ell)})
 \quad i \in \mathcal{S}\\\label{decen_svd_v}
v_{jh}^{(\ell+1)}&=\frac{1}{T}
\sum_{t=0}^{T-1}r_i(t)
A{{C}_{j}}({{\bm s}^{H}}(t){\bm u}_{h}^{(\ell)}) + \alpha {v}_{jh}^{(\ell)}\\\notag
&-\frac{1}{T}\sum_{m=1}^{h-1}\sum_{t=0}^{T-1}
{{\hat\sigma_{jm}}}^2
\hat{v}_{jm}
A{{C}_{j}}({\hat{{\bm u}}_m^{H}}{\bm u}_{h}^{(\ell)})
 \quad j \in \mathcal{R}
\end{align}
After a sufficient number of iterations $\ell ^\star$,  each eigenvector is normalized by:
\begin{align} 
	\scriptsize
\hat{u}_{ih}=\tfrac{u_{ih}^{\ell^\star}}{
\sqrt{A{{C}_{i}}\left({{({\bm u}_{h}^{\ell ^\star})}^{H}}{\bm u}_{h}^{\ell ^\star}\right)}},~
\hat{v}_{jh}=\tfrac{{v}_{jh}^{\ell^\star}}{\sqrt{A{{C}_{j}}\left({{({\bm v}_{h}^{\ell ^\star})}^{H}}{\bm v}_{h}^{\ell ^\star}\right)}\label{normal_svd}}.
\end{align}
The pseudo code is provided in Algorithm \ref{alg:2} and four remarks are in order \cite{wu2019decentralized}: 1) Algorithm \ref{alg:2} is sequential in a sense that iteration $h$ depends on the calculation results in previous $(h-1)$ iterations; 2) in each power iteration, ACG is evoked in the computation of equation (3)-(7) to obtain the $h$th eigenvectors and the corresponding singular values; 3) each ACG operation needs $K$ gossip iteration to reach a consensus among the neighboring nodes; 4) it is also worth mentioning that the sets $\mathcal{S}$ and $\mathcal{R}$ have to exchange the messages for calculating the  $A{{C}_{i}}({{\bm r}^{H}}(t){\bm v}_{h}^{(\ell)})$ and $A{{C}_{j}}({{\bm s}^{H}}(t)$ ${\bm u}_{h}^{(\ell)})$, unlike the distributed EVD process in \cite{wu2019decentralized}, where each set $\mathcal{R}$ or $\mathcal{S}$ operates its own calculation within their own network. Note that, we herein omit the algorithm for the power EVD, as it can be easily derived as a special case of power SVD.

%
\begin{algorithm}[t] 
\textbf{Input}:  Initialize for each agent an independent random vector, i.e.
${\bm u}_h^{(0)} \sim {\cal CN}({\bm 0}, {\bm I})$ for all $i \in \mathcal{S}$,  ${\bm v}_h^{(0)} \sim {\cal CN}({\bm 0}, {\bm I})$ for all $j \in \mathcal{R}$, where $h=1,2,\cdots,H$,. Set parameters $H$, $\ell^\star$ and $K$.\\
\textbf{for $h=1,2,\cdots,H$}
\par\setlength\parindent{2em} \textbf{for $\ell=0,1,\cdots,\ell^{\star}-1$}
\par\setlength\parindent{2em} $\bullet$ {Calculate each entry of ${\bm u}_{h}^{(\ell)}$ by \eqref{decen_svd_u} where
${{\hat\sigma_{im}}}^2$, $m=1,...,h-1$, is calculated by \eqref{decen_svd_u_eigenvalue}; each entry of ${\bm v}_{h}^{(\ell)}$ by \eqref{decen_svd_v} where ${{\hat\sigma_{jm}}}^2$, $m=1,...,h-1$, is calculated by \eqref{decen_svd_v_eigenvalue}.
}
\par\setlength\parindent{2em} \textbf{end for}\\
$\bullet$  Normalize ${\bm u}_{h}^{(\ell^\star)}$ and ${\bm v}_{h}^{(\ell^\star)}$ at each node by \eqref{normal_svd}.\\
\textbf{end for} \\
\textbf{Output}: 
For $h=1,2,\cdots,H$: ${{\bm u}_{h}}\triangleq {\hat{{\bm u}}}_{h}$ and ${{\bm v}_{h}}\triangleq {\hat{{\bm v}}}_{h}$ .
\caption{The Sequential Distributed Power SVD}\label{alg:2}
\end{algorithm}

\section{The parallel power method}\label{sec.decentr}
The power SVD method is based on a sequential process, wherein at each power iteration, only one eigenvector/eigenvalue is calculated via neighborhood communication which is initialized by a shaking-hand process.
A more efficient way is to exchange more information in each launch of shaking-hand by updating all the eigenvectors simultaneously via the gossip algorithm. 
This results in the following parallel power SVD. 
The idea of parallelization is to use the inexact intermediate  results of the top $(h-1)$ eigenvalues/eigenvectors to calculate the $h$th one. That is, at each iteration all eigenvalues and eigenvectors are calculated simultaneously despite that the results may not be precise.
To proceed, we first parallelize the way of obtaining the top $m$ ($m < h$) singular values: 

\begin{align}
{{\hat\sigma_{im}}}^2=&\frac 1 {T}
\sum_{t=0}^{T-1} \frac{{{s}_{i}}(t)A{{C}_{i}}({{\bm r}^{H}}(t){\bm v}_{m}^{(\ell)})}{{{\left( u_{im}^{(\ell)} \right)}^{*}}\sqrt{\tfrac{A{{C}_{i}}\left( {{\left( {\bm v}_{m}^{(\ell)} \right)}^{H}}{\bm v}_{m}^{(\ell)} \right)}{A{{C}_{i}}\left( {{\left( {\bm u}_{m}^{(\ell)} \right)}^{H}}{\bm u}_{m}^{(\ell)} \right)}}} , \quad \forall i \in \mathcal{S} \label{para_svd_sigi}\\
{{\hat\sigma_{jm}}}^2=&\frac 1 {T}
\sum_{t=0}^{T-1} \frac{{{r}_{j}}(t)A{{C}_{j}}({{\bm s}^{H}}(t){\bm u}_{m}^{(\ell)})}{{{\left( v_{jm}^{(\ell)} \right)}^{*}}\sqrt{\tfrac{A{{C}_{j}}\left( {{\left( {\bm u}_{m}^{(\ell)} \right)}^{H}}{\bm u}_{m}^{(\ell)} \right)}{A{{C}_{j}}\left( {{\left( {\bm v}_{m}^{(\ell)} \right)}^{H}}{\bm v}_{m}^{(\ell)} \right)}}}. \quad \forall j \in \mathcal{R} \label{para_svd_sigj}
\end{align}
Then each entry of the left and right eigenvectors can be obtained by:
\begin{align}
&u_{ih}^{(\ell+1)}=\frac{1}{T}
\sum_{t=0}^{T-1}s_i(t)
AC_i({\bm{r}^H(t)\bm{v}_{h}^{(\ell)}}) + \alpha {u}_{ih}^{(\ell)}\label{para_svd_u} \quad \forall i \in \mathcal{S} \notag\\
&-\tfrac{1}{T}\sum_{m=1}^{h-1}\sum_{t=0}^{T-1}
 \tfrac{{{\hat\sigma_{im}}}^2
u_{im}^{(\ell)}
AC_i\left(\left(\bm{v}_m^{(\ell)}\right)^H \bm{v}_h^{(\ell)}\right)} {\sqrt{AC_{i}\left({{\left(\bm {u}_{m}^{(\ell)}\right)}^{H}}{\bm u}_{m}^{(\ell)}\right)AC_{i}\left({{\left({\bm v}_{m}^{(\ell)}\right)}^{H}}{\bm v}_{m}^{(\ell)}\right)}} \\ 
&v_{jh}^{(\ell+1)}=\frac{1}{T}
\sum_{t=0}^{T-1}r_j(t)
AC_j(\bm{s}^H(t)\bm{u}_{h}^{(\ell)}) + \alpha {v}_{jh}^{(\ell)}\label{para_svd_v}\quad \forall j \in \mathcal{R}\notag\\
&-\tfrac{1}{T}\sum_{m=1}^{h-1}\sum_{t=0}^{T-1}    
\tfrac{{{\hat\sigma_{jm}}}^2
v_{jm}^{(\ell)}
AC_j\left(\left(\bm{u}_m^{(\ell)}\right)^H \bm{u}_h^{(\ell)}\right)} {\sqrt{AC_{j}\left({{\left(\bm {u}_{m}^{(\ell)}\right)}^{H}}{\bm u}_{m}^{(\ell)}\right)AC_{j}\left({{\left({\bm v}_{m}^{(\ell)}\right)}^{H}}{\bm v}_{m}^{(\ell)}\right)}}. 
\end{align}}
The parallel power SVD is summarized in Algorithm \ref{alg:4}. 
\begin{algorithm}[t] 
\textbf{Input}:  Initialize for each agent an independent random vector, i.e.
${\bm u}_h^{(0)} \sim {\cal CN}({\bm 0}, {\bm I})$ for all $i \in \mathcal{S}$,  ${\bm v}_h^{(0)} \sim {\cal CN}({\bm 0}, {\bm I})$ for all $j \in \mathcal{R}$, where $h=1,2,\cdots,H$,. Set parameters $H$, $\ell^\star$ and $K$\\
\textbf{for $\ell=0,1,\cdots,\ell^\star-1$}
\par\setlength\parindent{2em} \textbf{for $h=1,2,\cdots,H$}
\par\setlength\parindent{2em} $\bullet$ {Calculate each entry of ${\bm u}_{h}^{(\ell)}$ by \eqref{para_svd_u} where ${{\hat\sigma_{im}}}^2(m=1,...,h-1)$ is calculated by \eqref{para_svd_sigi} and each entry of ${\bm v}_{h}^{(\ell)}$ by \eqref{para_svd_v} where  ${{\hat\sigma_{jm}}}^2(m=1,...,h-1)$ is calculated by \eqref{para_svd_sigj}
}
\par\setlength\parindent{2em} \textbf{end for}\\
\textbf{end for} \\
$\bullet$  Normalize ${\bm u}_{h}^{\ell^\star}$ and ${\bm v}_{h}^{\ell^\star} (h=1,2,\cdots,H)$ at each node by \eqref{normal_svd}.\\
\textbf{Output}: For $h=1,2,\cdots,H$: ${{\bm u}_{h}}\triangleq {\hat{\bm u}_{h}}$ and ${{\bm v}_{h}}\triangleq {\hat{\bm v}_{h}}$.
\caption{The Parallel Distributed Power SVD}\label{alg:4}
\end{algorithm}

{\it Remark:}
To compare the times of node-to-node shaking-hand between the sequential and parallel power method, assume that we aim at finding $H$ eigenvectors for the covariance matrix ${\bm R_{ss}}$ (${\bm R_{rr}}$), each of which needs $\ell^{\star}$ times of power iteration and one time of normalization, wherein $K$ times of gossip iterations is needed to reach a consensus. For EVD power method is $HK(\ell^{\star}+1)$ times of shaking-hand for the sequential case and $K(\ell^{\star}+1)$ for the parallel case.

Similarly,  supposing that $\mathcal{S}$ runs gossip algorithm $K_1$ times and $\mathcal{R}$ runs gossip algorithm $K_2$ times, the times of shaking-hand for the sequential power SVD is $H(K_{1}+K_{2})(\ell^{\star}+1)$, while for the parallel power SVD is reduced to $(K_{1}+K_{2})(\ell^{\star}+1)$. 
Obviously, the times of shaking-hand of the parallel method is about $1/H$ times of that of the sequential method. The reduction comes from that we transmit more information at every node-to-node shaking-hand.

\section{SIMULATIONS}
In this section, we compare the performance of the proposed parallel DPM method with its sequential counterparts and the distributed LA method proposed in \cite{penna2014decentralized}. As distributed LA can only calculate the eigenvalues for a Hermitian matrix, we consider the normalized mean square error (NMSE) for the eigenvalue computation as a performance measurement for the power EVD:
\[
	{\text {NMSE}_{\text {EVD}}} =  {\sum\limits_{i=1}^{p} {\left| \left| \lambda_{true}^i - \lambda_{est.}^i  \right| \right|^2}} / {\sum \limits_{i=1}^p {\left( \lambda_{true}^i \right)^2}},
\]
while for the comparison of the sequential and parallel DPM, we define the NMSE of eigenvectors for the power SVD:
\[
	{\text {NMSE}_{\text {SVD}}}=\tfrac{1}{2}\sum\limits_{i=1}^{p}{\tfrac{\left\| \left. {{\bm u}_{est.}^i}-{{\bm u}_{true}^i} \right\| \right._{2}^{2}}{\left\| \left. {{\bm u}_{true}^i} \right\| \right._{2}^{2}}+}\tfrac{\left\| \left. {{\bm v}_{est.}^i}-{{\bm u}_{true}^i} \right\| \right._{2}^{2}}{\left\| \left. {{\bm v}_{true}^i} \right\| \right._{2}^{2}}.
\]
It is also worth noting that the iteration time $M$ in distributed LA and the number of calculated eigenvalues $H$ are often tied together, i.e., when calculating top $H$ eigenvalues of matrix ${\bm R}_{ss} $, one should set  $H \le M \le {\rm rank}({\bm R}_{ss})$ .   
 
To set up the experiment,  we consider the scenario of passive radar, where signal ${\bm s}(t)$ and ${\bm r}(t)$ are generated by:
$$
\bm{s}(t)={\bm H}_s{\bm{\theta}}_{s}(t)+{\bm{w}}_s(t),~
\bm{r}(t)={\bm H}_r{\bm{\theta}}_{r}(t)+{\bm{w}}_r(t)
$$
where $\bm{\theta}_{i}(t) =[\theta_{1}(t),\theta_{2}(t), \ldots \theta_{|{\cal W}_i|}(t)]^T\sim {\mathcal CN}({\bm 0}, \alpha_{i} {\bm I})$ ($i = s,r$) represent the vector of the signal generated at $\mathcal{S}$ and $\mathcal{R}$ by the transmitters, ${\bm{w}}_s(t), {\bm{w}}_r(t) \sim {\mathcal CN}({\bm 0}, {\bm I})$ are the standard Gaussian noise, and ${\bm H}_s, {\bm H}_r  \sim {\mathcal CN}({\bm 0}, {\bm I})$ represent the channels. Herein, we have ${\rm rank}({\bm R}_{ss})=|{\cal W}_i|$. We therefore consider a network with $10$ nodes for set $\mathcal{S}$ and $12$ nodes for set $\mathcal{R}$. The graph underlying $\mathcal{S}$ is a small-world graph with degree $4$ and rewiring probability $0.2$, underlying $\mathcal{R}$ is a small-world graph with degree $6$ and rewiring probability $0.2$. There are also connections between $\mathcal{S}$ and $\mathcal{R}$ which could be wired and wireless so that necessary information could be exchanged for SVD. Within $\mathcal{S}$ or $\mathcal{R}$, we follow the reference \cite{xiao2004fast} to set the 
$
{\bm W}=I-\frac{2}{{{\lambda }_{1}}(\bm L)-{{\lambda }_{N-1}}(\bm L)}{\bm L}
$
where ${\bm L}$ is the graph Laplacian matrix and ${{\lambda }_{i}}({\bm L})$ is the $i$th largest eigenvalue of ${\bm L}$. With the above parameter settings, we consider a PCA trail as the calculation of the covariance (correlation) matrix of $T$ = 500 samples of signal.  In total $500$ PCA trails are calculated to get an averaged NMSE performance of calculating top $H$ eigenvalues/eigenvectors. The ACG iteration is denoted by $K$.

\begin{figure}[t!]
	\centering
	\includegraphics[width = 7.2cm]{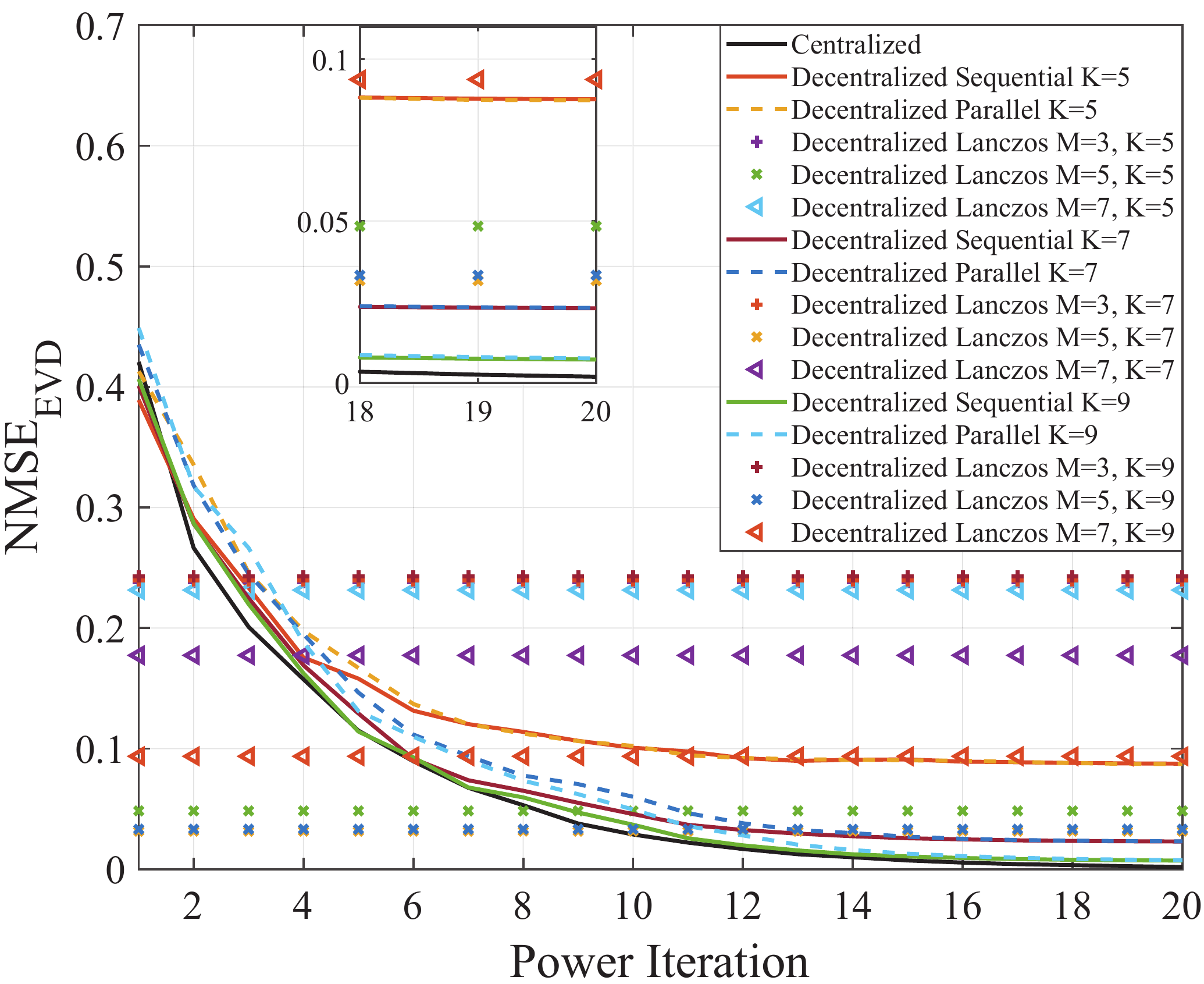}
	\caption{\scriptsize NMSE curves for top three eigenvalues  of $\bm {R_{ss}}$ via EVD.}
	\label{Fig:3}
\end{figure}
\begin{figure}[t!]
	\centering
	\includegraphics[width = 7.2cm]{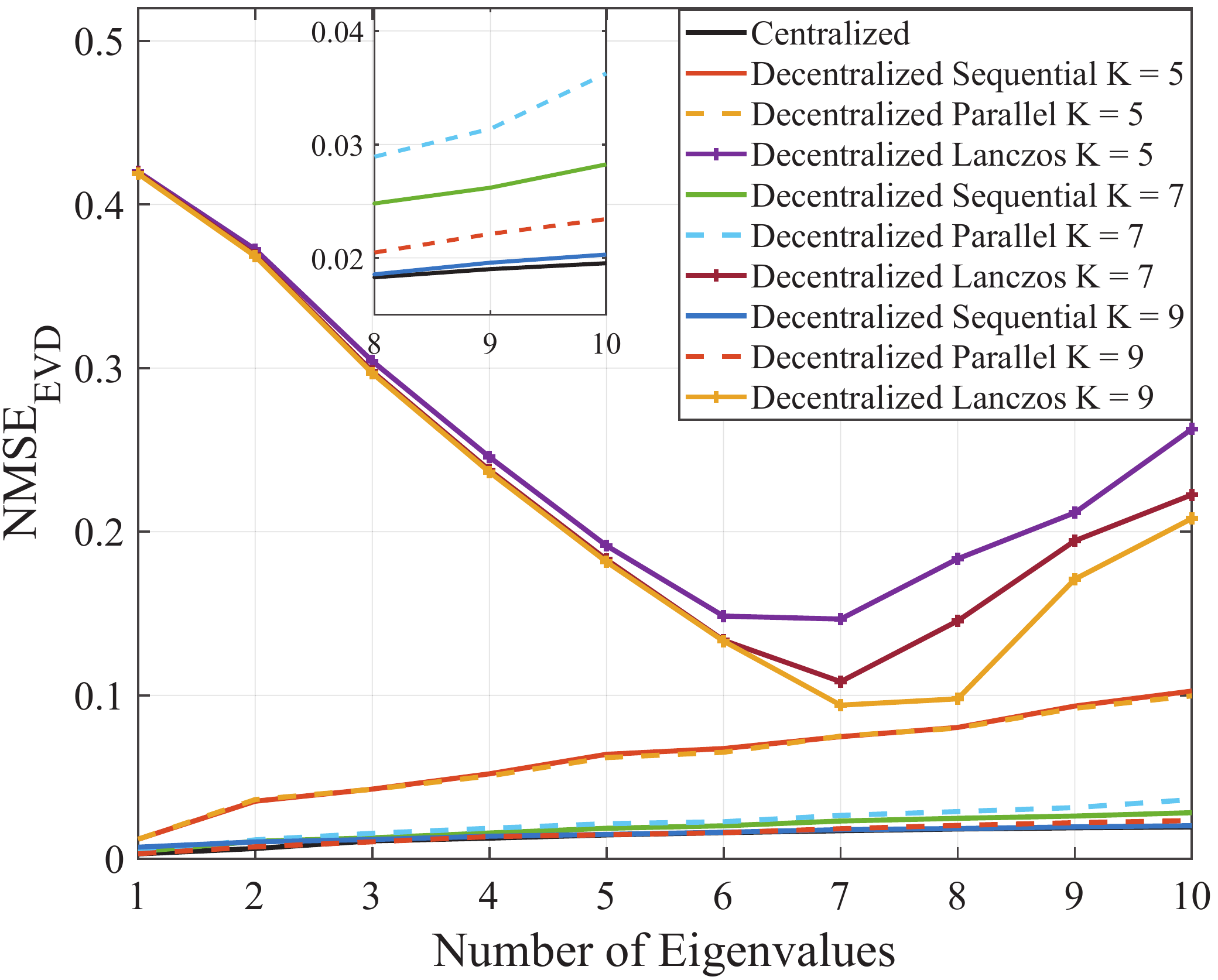}
	\caption{\scriptsize NMSE curves for eigenvalues of $\bm {R_{ss}}$ via EVD.}
	\label{Fig:4}
\end{figure}
\begin{figure}[t!]
	\centering
	\includegraphics[width =  7.2cm]{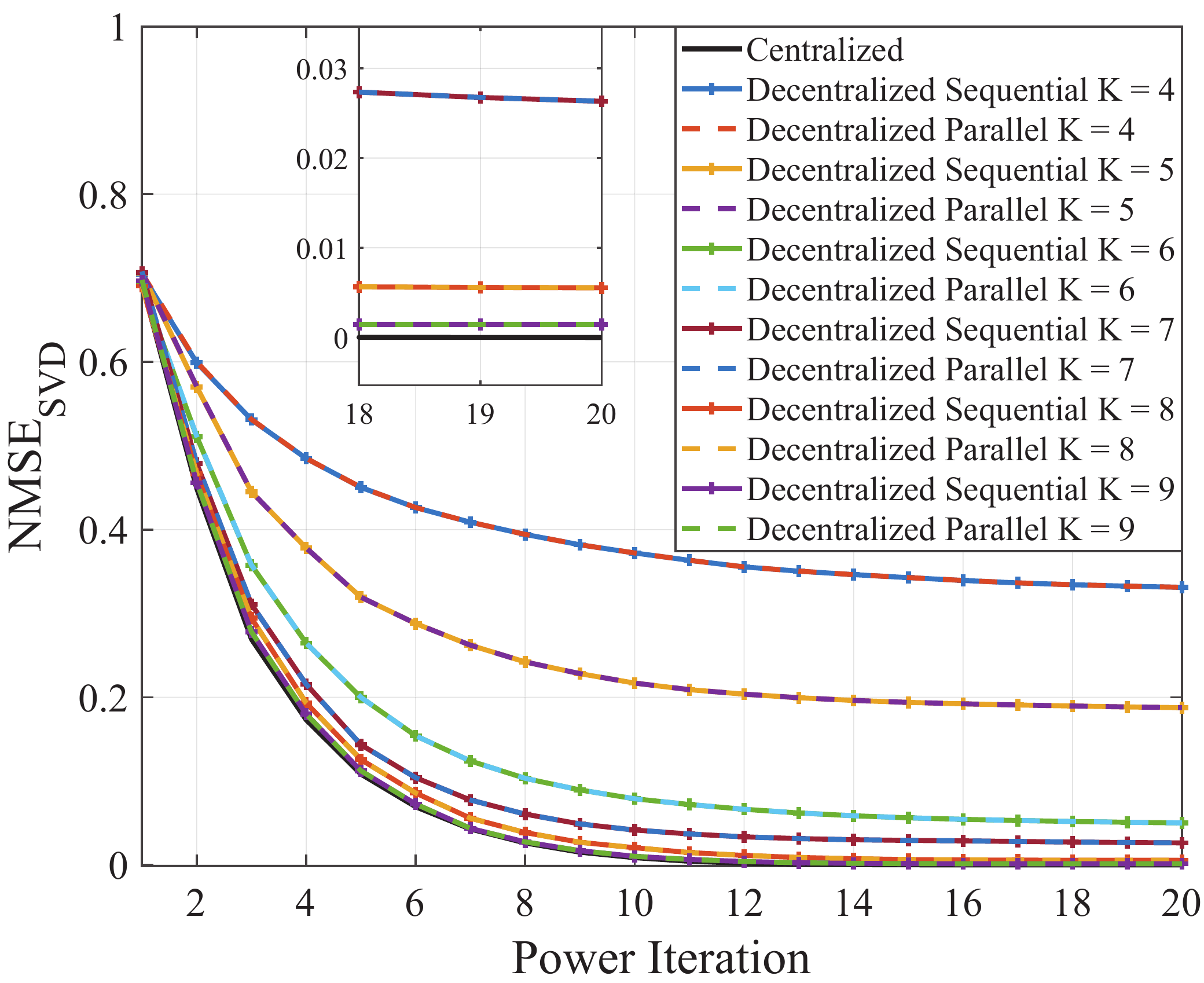}
	\caption{\scriptsize NMSE curves for top two left and right eigenvectors of $\bm {R_{sr}}$ via SVD.}
	\label{Fig:5}
\end{figure}

In Fig.~\ref{Fig:3}, we compare NMSE curves for top $H=3$ eigenvalues of the covariance matrix $\bm {R_{ss}}$ via the power EVD method and the LA method. Specifically, we set $|{\cal W}_s|=10$ and vary the number of power iteration from $\ell^\star=1$ to $\ell^\star=20$, the number of LA iteration from $M=3$ to $M=7$. From the figure we see that when the number of power iteration is not sufficient to ensure its convergence, both sequential and parallel distributed PMs are inferior to distributed LA, while once $\ell^\star$ is sufficient, they both outperform distributed LA and approach to the centralized PM. We also observe that the performance of distributed LA improves from $M=3$ to $M=5$, while deteriorates from $M=5$ to $M=7$. This phenomenon is due to the reason that when the number of LA iteration $M$ is close to the rank of the matrix, i.e., $M \approx$ rank(${\bm R}_{ss}$),  the orthogonality of the Lanczos vectors might be destructed, thereby leading repeated roots of eigenvalue in the calculation \cite{golub2013matrix}, which are difficult to differentiate in a decentralized case. 
Moreover, as $K$ increases, NMSEs for the proposed methods decrease and they will approach to the centralized EVD when $K$ is sufficiently large. Moreover, the sequential EVD and the parallel EVD have a comparable NMSE, given that parallel EVD only requires $1/H$ times of node-to-node shaking-hand of that in the sequential case.
In Fig.~\ref{Fig:4}, we set $|{\cal W}_s|=10$, $\ell^\star=20$ and vary the number of calculated eigenvalues from $H=1$ to $H=10$ while letting $M=H$. We can see that the performance of sequential and parallel PM are comparable, while distributed LA first improves and then deteriorates with $M$ and $H$. This is consistent with what we have observed in Fig.~\ref{Fig:3}.
In Fig.~\ref{Fig:5}, we further compare the sequential and parallel power methods in the SVD case with $|{\cal W}_r|=|{\cal W}_s|=2$ and $\ell^\star=20$.  The curves in Fig.~\ref{Fig:5} show a similar trend as the EVD case, where the sequential and parallel SVD have comparable NMSEs and  both approach to the centralized PM when the gossip iteration $K$ is sufficiently large. 

\section{Conclusions}
We have introduced an improved distributed power method, which transfers the traditional sequential power iteration to a parallel one in the process of the decentralized power method. The basic idea is to exchange more information in every launch of shaking-hand in the gossip algorithm by allowing all eigenvectors to update simultaneously in each power iteration. Simulation results show that the proposed parallel method exhibits a similar error performance as its sequential counterparts for both EVD and SVD calculations, while the communication cost of the parallel method is about $1/H$ times of that for the sequential counterpart, given that $H$ is the number of eigenvectors to compute.

\bibliographystyle{IEEEtran}
\bibliography{ref_pr}
\end{document}